# Ultrafast humidity sensor based on liquid phase exfoliated graphene

Stevan Andrić[1], Tijana Tomašević-Ilić[2], Marko V. Bošković[1], Milija Sarajlić[1], Dana Vasiljević-Radović[1], Milče. M. Smiljanić[1] and Marko Spasenović[1*]

[1]Center for Microelectronic Technologies, Institute of Chemistry, Technology and Metallurgy, University of Belgrade, Njegoševa 12, 11000 Belgrade, Serbia

[2]Institute of Physics Belgrade, University of Belgrade, Pregrevica 118, 11080 Belgrade, Serbia

E-mail: spasenovic@nanosys.ihtm.bg.ac.rs

**Abstract**

Humidity sensing is important to a variety of technologies and industries, ranging from environmental and industrial monitoring to medical applications. Although humidity sensors abound, few available solutions are thin, transparent, compatible with large-area sensor production and flexible, and almost none are fast enough to perform human respiration monitoring through breath detection or real-time finger proximity monitoring via skin humidity sensing. This work describes chemiresistive graphene-based humidity sensors produced in few steps with facile liquid phase exfoliation (LPE) followed by Langmuir-Blodgett assembly that enables active areas of practically any size. The graphene sensors provide a unique mix of performance parameters, exhibiting resistance changes up to 10% with varying humidity, linear performance over relative humidity (RH) levels between 8% and 95%, weak response to other constituents of air, flexibility, transparency of nearly 80%, and response times of 30 ms. The fast response to humidity is shown to be useful for respiration monitoring and real-time finger proximity detection, with potential applications in flexible touchless interactive panels.

# 1. Introduction

Humidity monitoring is essential for numerous applications across industries, such as environmental and industrial monitoring, and healthcare [1,2]. Aside from traditional uses in monitoring atmospheric and room conditions, technological progress keeps enabling new uses for humidity sensors. In healthcare, humidity sensing could be used for human respiration monitoring [3,4] due to the high level of water vapor in breath. In electronics and robotics industries, humidity clouds near human skin could be used to detect finger position for touchless control interfaces [5,6]. However, humidity sensors made of established materials such as metal/polymer composites have temporal response rates that are far too slow for these applications, in the range 5–50 s [2]. Emerging composite and nanomaterials such as ZnO and Pd-$SnO_2$ also suffer from the same debilitating disadvantage [7,8]. More recently, graphene and graphene oxide have emerged as a promising material for fast humidity sensing, with response times ranging from 10's of milliseconds to a few seconds, depending on the method of production [9–15].

Compared to many other materials, graphene has the added benefits of being thin, flexible, and transparent, enabling applications in wearable and flexible electronics. Nevertheless, graphene humidity sensors to date have been made from graphene that is either industrially irrelevant (mechanically exfoliated), expensive (chemical vapor deposited, CVD) or made with several complex chemistry steps, such as with reduction of graphene oxide.

Here, we demonstrate fast humidity sensors made with an inexpensive and facile production method that is compatible with large-area sensor production. The active sensing area is made from liquid-phase exfoliated graphene [16] that requires only a single ultrasonic processing step. The humidity response times of our sensors are as low as ~30 ms, which allows us to show real-time breath monitoring and finger proximity detection as exemplary applications of ultrafast humidity sensing. We demonstrate sensing capability on three different substrates, including flexible transparent ones. The sensors fare very well against standard gas sensor performance metrics, such as insensitivity to other components of air and response time [17].

## 2. Material and methods

*2.1 Fabrication of graphene films and humidity sensors*

The graphene dispersion was produced by dissolving graphite powder (Sigma Aldrich, product no. 332461) at a concentration of 18 mg ml$^{-1}$ in N-Methly-2-pyrrolidone (NMP) (Sigma Aldrich, product no. 328634). The dispersion was sonicated in a low energy ultrasonic bath for 14 h. After sonication the dispersion was centrifuged at 3000 rpm for 60 min in order to separate non-exfoliated graphite flakes, which remain in the precipitate, and the exfoliated graphene flakes which are dispersed in the supernatant. A small volume of the supernatant is added to deionized water (18 MΩ cm$^{-1}$) resulting in self-assembly of graphene nanoplatelets into a thin film on the water/air interface. The thin film is deposited on a pre-immersed substrate of choice following the Langmuir-Blodgett method [16,18–20].

Our film, made of graphene that is exfoliated in the liquid phase and assembled with the Langmuir-Blodgett method, consists of graphene nanoplatelets in contact with each other, as shown in the atomic force micrograph in Supporting Information (Figure S1). Although nanoplatelets conform to a distribution of thicknesses, the average thickness of the film is ~10 layers of graphene (3.4 nm), as measured with UV-VIS absorption spectroscopy and shown in Supporting Information (Figure S2). For well-defined channel geometry and accurate sheet resistance measurements, we use a SiO$_2$/Si substrate with four microfabricated metal contacts (Figure 1a). The substrate is a 380 μm thick n-doped Si wafer with a thermally grown 800 nm thick SiO$_2$ insulating layer. We deposit a layer of chromium (20 nm) and a layer of gold (100 nm) with radio-frequency cathode sputtering. Subsequently the layers of chromium and gold are coated with 0.5 μm of photoresist (AZ-1505) that is subjected to direct laser writing (LW405, MicroTech, Italy) [21] to pattern the contacts. The chromium and gold are removed with a solution of potassium iodide. The wafer is diced into 3 x 3 mm chips, each chip containing a set of four metal contacts. After film deposition, the chips are mounted to TO-8 housing.

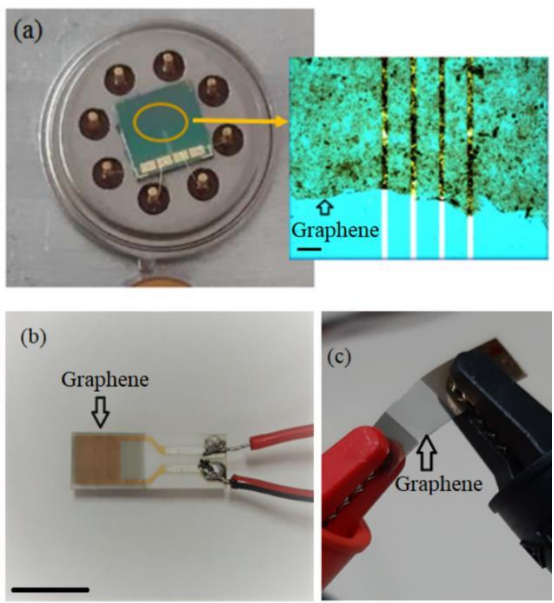

**Figure 1.** (a) Graphene humidity sensors on different substrates. (a) Optical image of a sensor with four contacts and the graphene sensor active area. Inset: optical micrograph of graphene film on contacts. The scale bar is 50 μm. (b) Optical image of the graphene film on the ceramic commercial substrate with interdigitated electrodes. The scale bar is 1 cm. (c) Optical image of graphene film on PET with macroscopic gold contacts.

The inset of Figure 1a depicts an optical micrograph of the graphene film deposited on the contacts. Darker spots indicate remaining unexfoliated or thick graphite material. The film area between contacts has dimensions of ~1500 μm x 190 μm. Taking into account film geometry yields sheet resistance of 3-7 k$\Omega$ sq$^{-1}$ for our films, the smallest values reported for post-processing-free single-deposition Langmuir-Blodgett graphene films to date [16,18,20,22]. Such small sheet resistance is a result of fabrication process streamlining and careful four-terminal resistance measurements. For obtaining the largest sensing area and highest signal-to-noise ratio, we employ a commercial ceramic substrate with pre-made interdigitated electrodes (DropSens IDEAU200), Figure 1b. This substrate is easy to handle, versatile, low-cost, and easily connected to macroscopic wires by soldering, and the active sensing area is ~ 15 mm$^2$. Finally, for flexible transparent humidity sensors we employ a polyethylene terephthalate (PET) substrate with macroscopic gold contacts thermally evaporated over a shadow mask, as in Figure 1c. The size of the active area on this substrate is 5 x 5 mm. The optical transmittance at a wavelength of 660 nm is 77%, as shown in Figure S2 of Supporting Information.

*2.2 Humidity sensing*

We perform humidity sensing in a homebuilt humidity chamber made of polytetrafluoroethylene (PTFE), as depicted in Figure 2. The chamber is equipped with separate valves for injection of water vapor and another gas, a gas outlet valve, and auxiliary connectors. Our sensor in its TO-8 housing is integrated into a custom-made PTFE plug that we insert into a matching slot in the chamber, next to a reference humidity sensor (Honeywell HIH-4000-001). A thermocouple is placed near the sensors to measure the local temperature. All measurements were performed at room temperature (21-23 °C). We apply a current of 10 µA between the outer electrode pair and measure the induced voltage across the inner electrode pair to obtain the resistance. The voltage was measured with a Keysight 34461a DMM. In the cases of the commercial ceramic substrate and the PET substrate we monitored two-terminal resistance with the same DMM operated in ohmmeter mode. Relative humidity was decreased to 8% by purging the sealed chamber with nitrogen ($N_2$) gas. Water vapor is produced with a commercial room humidifier and injected into the chamber at a constant flow rate. Once a desired humidity level is reached the water vapor inlet valve is manually closed.

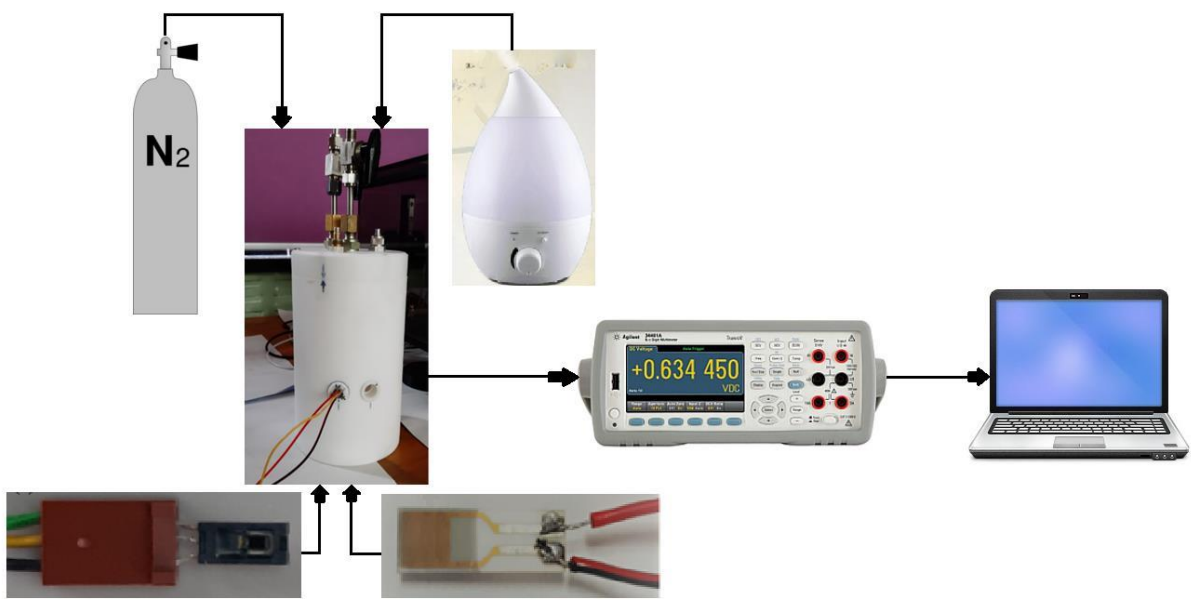

**Figure 2**. Experimental setup for measuring humidity response. A homebuilt humidity chamber is designed with slots for the graphene sensor and a reference sensor. It has inlets for water vapor from a room humidifier and for purging with $N_2$ gas. The sensors are connected to measurement electronics.

To measure sensor speed below the limit imposed by humidity chamber filling time, we placed the sensors on a table in free space at RH ~40%. A nitrogen gun was used to induce nitrogen flow across the sensor surface, quickly drying the active area while we monitored the sensor response and recovery times [11].

*2.3 Respiration monitoring and touchless sensing*

Respiration monitoring and touchless sensing were performed at atmospheric conditions, at a temperature of ~25 °C and RH ~40%. The graphene sensors were placed on a table and connected to measurement electronics as described above, and a volunteer proceeded to breathe onto the sensor surface or bring a finger nearby the sensor. All three types of substrates were tested under the same conditions.

## 3. Results and discussion

*3.1 Humidity sensing*

The humidity sensing performance of our devices was tested by monitoring device resistance while controlling relative humidity in the humidity chamber. Figure 3a depicts three cycles of a humidity ramp in the test chamber and corresponding measurements with our graphene sensor on the ceramic substrate with interdigitated electrodes (black) and the reference sensor (purple). During each cycle the humidity in the chamber is increased from ~8% to 95% and then decreased to ~8%. The resistance of our sensor rises with humidity, as was shown earlier for other films that consist of conducting $NbS_2$ nanoplatelets [23]. The increase in resistance is attributed to water adsorption at nanoplatelet edges and between the platelets, both of which disrupt electrical conduction paths of the film. In the case of graphene, other mechanisms could also play a role, such as electron donation from graphene to water and disruption of molecular symmetries of graphene by the water molecules [24]. For a clear perception of the sensitivity of our sensor, we plot a second vertical axis next to the resistance axis that depicts the percent change of resistance, described by $S = 100 \cdot \frac{\Delta R}{R_0}$, where $R_0$ is the initial resistance value and $\Delta R$ is the difference between the given and the initial resistance value. The resistance of the graphene sensor changes by 5% when changing humidity from 8% to 95%. The sensitivity is thus higher than reported for CVD graphene [10,11], which is more costly than LPE graphene, and higher than for industrially irrelevant mechanically exfoliated graphene [25]. Sensor repeatability indicates that the likely dominant mechanism is physisorption of water molecules, with little chemisorption. There is a small baseline drift that occurs when the sensor rests in air which is visible on Figure 3a. A similar drift was observed in mechanically exfoliated bilayer graphene [11]. Although such a drift could potentially be detrimental to the practical use of graphene-based gas sensors, we found that the drift is fully reversible with heating to temperatures around 150 °C that could easily be achieved with an on-chip integrated heater [26].

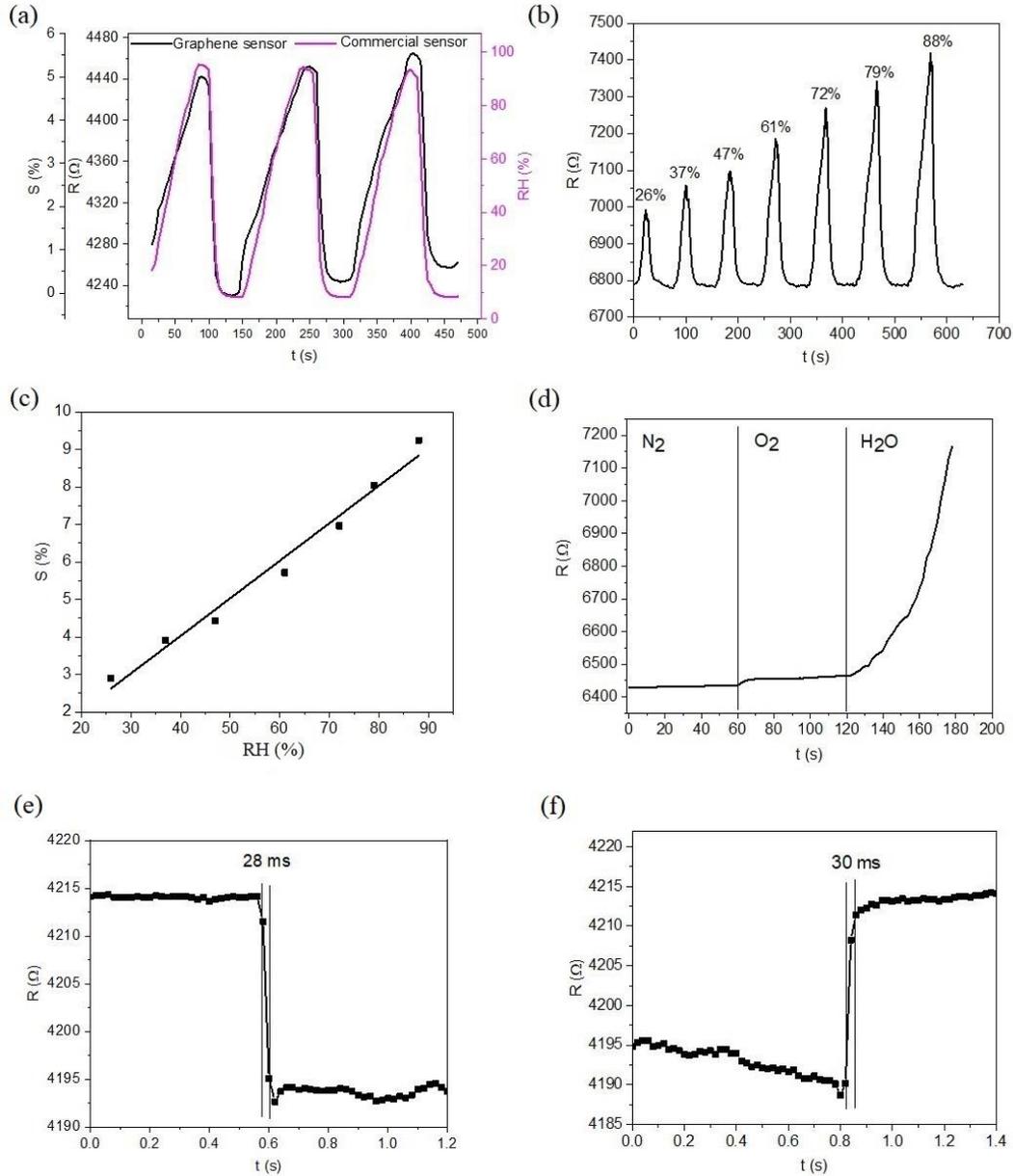

**Figure 3.** Response to humidity and other constituents of air. Measurements are conducted on the ceramic substrate with interdigitated electrodes. (a) Graphene sensor response measured in conjunction with the reference sensor response over three cycles of ramping RH from ~8% to 95%. (b) Graphene sensor response to repeat stepwise increase of relative humidity, from 26% to 88%, in time. (c) Peak sensor response as function of maximum RH. (d) Sensor response to nitrogen, oxygen and water

vapor over 60 seconds each. (e) Response time of graphene sensor when an $N_2$ gun is used to flush the device. (f) Recovery time of graphene sensor after flushing with $N_2$ gun.

Figure 3b depicts the response of the graphene sensor over several cycles with different maximum RH. In this image the baseline drift has been corrected for by linear subtraction. The raw data that includes the drift is provided in Supporting Information (Figure S3). The relative change in resistance, S, as a function of maximum humidity is shown in Figure 3c. The sensor response is clearly linear with humidity (r =0.981), which indicates potential for applications in diverse conditions. In the case of our graphene sensors on $Si/SiO_2$ substrates, the linearity is similar to that reported here, while the sensitivity is ~10 times smaller (see Supporting Information, Figure S4).

To confirm that our sensor reacts to water vapor and not to other constituent gases of air, we tested the response to nitrogen and oxygen. Figure 3d depicts the sensor response to nitrogen gas (injected into the chamber at t = 0 s), to oxygen gas replacing nitrogen (at t = 60 s), and finally to water vapor injected instead of oxygen (at t = 120 s). The sensor does not respond to $N_2$, reacts very little to $O_2$ and has a strong response to $H_2O$ gas, an effect that could be used to implement selectivity. Humidity was reduced to ~0% before starting the experiment.

Sensor response and recovery times cannot be measured in the humidity chamber due to the limited chamber-filling and chamber-flushing times, hence we proceed to measure the sensor response time by rapidly drying the sensor surface in ambient with a nitrogen gun and observing sensor dynamics. In Figure 3e we show nitrogen gun drying of the sample, with observed rapid recovery shown in Figure 3f. We set a 10% and a 90% change threshold for measuring rise and fall times. The sensor responds in 28 ms and recovers in 30 ms. Similar dynamics are observed in other samples on the same substrate. The obtained response is significantly faster than that reported earlier for single-layer and double-layer CVD graphene (~700 ms) [10,11] and is two orders of magnitude faster than the commercial reference sensor (Honeywell HIH6100 Series Datasheet). The response time in the case of our graphene sensors on $Si/SiO_2$ substrates is longer, in the range of 240 ms (see Supporting Information Figure S5).

*3.2 Respiration monitoring*

High-speed humidity sensors enable the monitoring of human respiration by breath detection. To test the usefulness of our sensors for respiration monitoring, we placed the sensors on a table and had a volunteer breathe near the sensor surface. Figure 4a depicts the response of a sensor on a ceramic substrate to breathing cycles in a fast, regular, and slow pattern. A recording of respiration detection is shown in Supporting Video 1. It is evident that the sensor responds to human respiration and can be used as a biometric detector of respiration rate. In real-life conditions outside the humidity chamber, the resistance changes by up to 20% during breathing, which offers an excellent on/off ratio that is useful for applications. The sensitivity to breath is larger than can be found in literature that describes other graphene-based sensors, which are also generally produced with more complex chemistry steps [27]. The high sensitivity of our sensor is likely due to the abundancy of reactive edge sites in the film consisting of interconnected nanoplatelets. The sensor on the $SiO_2$/Si substrate also responds to breath, although with a smaller sensitivity (see Supporting Information Figure S6).

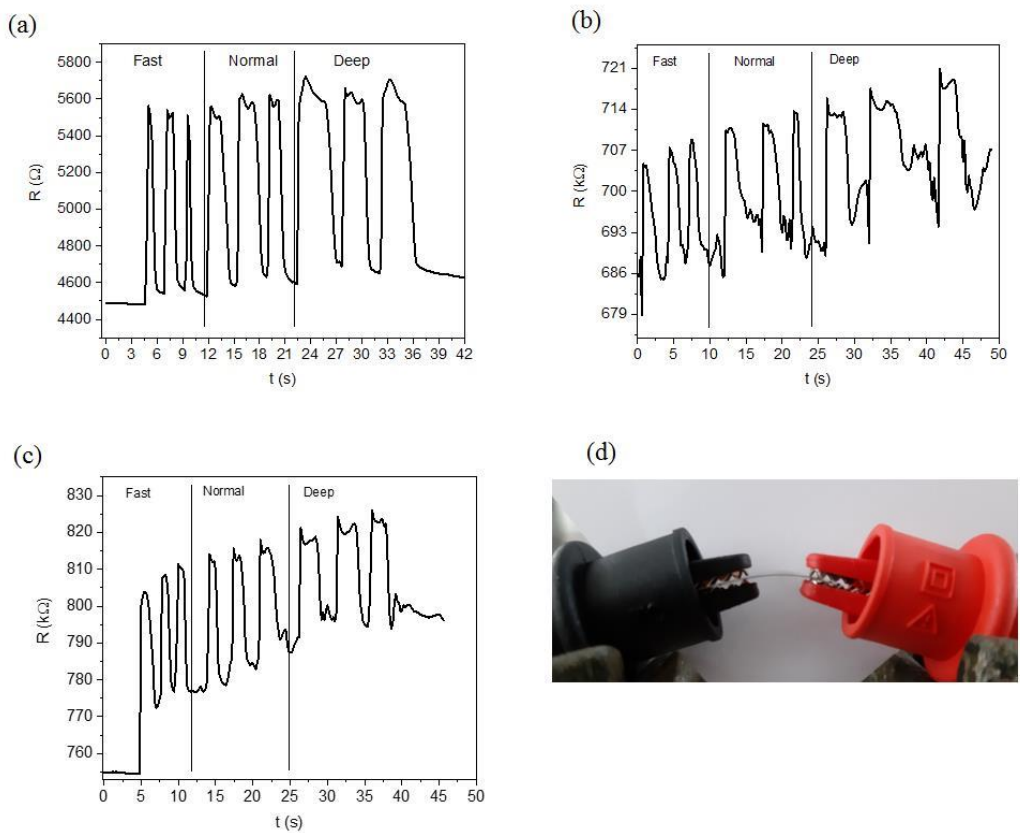

**Figure 4.** Respiration monitoring. (a) Monitoring fast, normal and deep breathing with graphene sensor on ceramic substrate. (b) Monitoring breathing rate on flat PET substrate, and (c) on PET that is bent at an angle of 10 degrees, as seen in (d).

*3.3 Transparent flexible sensors*

For certain applications, such as for monitoring the respiration rate of first responders or medical patients via sensors attached to transparent masks, or for transparent touchless control panels, it would be advantageous to have the sensor on a transparent substrate. Results of respiration monitoring of our large-area sensors on PET are shown in Figure 4b. Qualitatively, the performance is similar as on the rigid substrate. Quantitatively, the response is an order of magnitude weaker on PET, with a more pronounced background drift. The speed of sensor response on a PET substrate is ~20 ms, as shown in Supporting Information (Figure S7). We purposely made the sensor on a flexible substrate to demonstrate compatibility with flexible electronics and wearable technology. Flexing the substrate is not detrimental to the humidity sensing performance of the sensor, and results in similar response to breath, as seen in Figure 4c. The data shown in Figure4c was taken for a sensor bent with a curvature of 10 degrees, as in Figure 4d. The method of measuring curvature is presented in Supporting Information (Figure S8).

*3.4 Finger proximity detection*

Our devices have an interesting application in proximity sensing, as part of positioning interfaces for touchless screens and applications in robotics [5]. It is well known that human skin emits a cloud of moisture that decays over a distance of ~1 cm, an effect that has been proposed as a working mechanism for positioning interfaces that detect the presence of a human finger [6]. However, practical realization of such interfaces has been elusive, primarily due to the low speed of the materials considered thus far. Devices based on $VS_2$ [6] and graphene oxide [28–30] have response and recovery times ranging from 1 second to more than 20 seconds, which causes a delay in finger position detection that is impractical. Figure 5a depicts an optical image of the proximity detection experiment. A finger is held at a specific distance from the device as the resistance is measured. Figure 5b depicts the distance-dependent device response to the presence of a fingertip. The sensor responds at a finger distance of 10 mm and resistance significantly increases for smaller distances. To demonstrate ultrafast performance of our proximity sensor, the volunteer swipes his finger above the device at different distances (Figure 5c). It is clear that the device responds to finger motion in real time, enabling practical development of novel man-machine interactive systems. Real-time finger proximity detection is demonstrated in Supporting Video 2. We show that the response to human finger proximity is due to humidity and not a capacitive effect by testing the sensor response to the presence of metal tweezers and a finger covered with a rubber glove (Figure 5d).

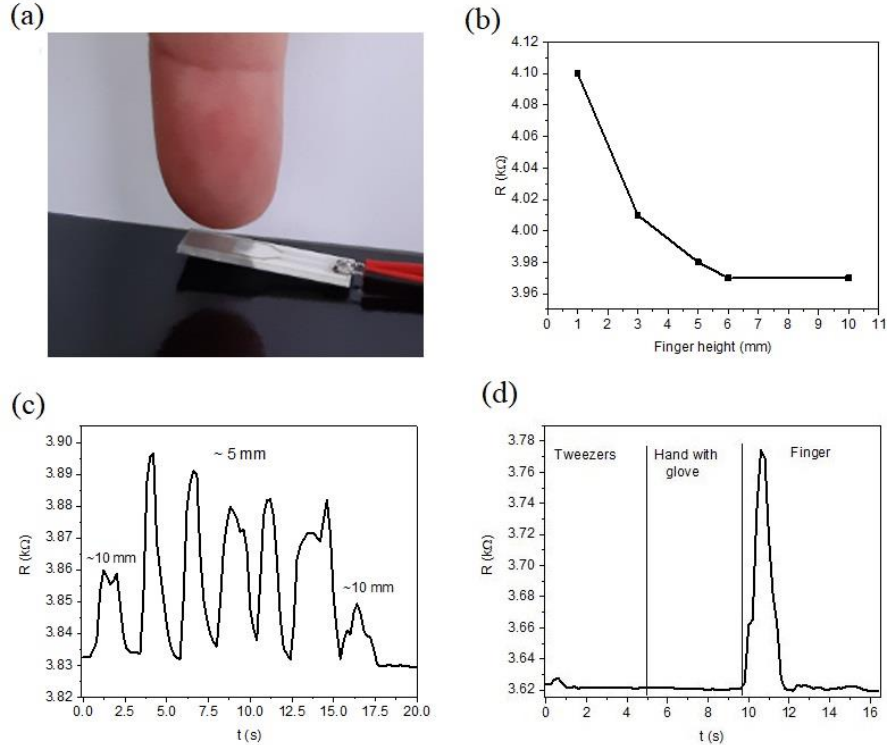

**Figure 5.** Finger proximity detection. (a) Optical image of the touchless proximity sensing experiment. (b) Resistance as a function of fingertip distance to device. (c) Demonstration of ultrafast proximity detection as a finger is swept across the device at different distances. (d) Response of touchless sensor to metallic tweezers and hand with glove.

## 4. Conclusions

We have demonstrated humidity sensors based on graphene that are sensitive, thin, flexible, nearly 80% transparent, only weakly reactive to other constituents of air, and fast enough to be used for advanced applications such as respiration rate monitoring and finger proximity detection. The principles of operation shown here, combined with the ease of manufacture of the sensors, indicate strong technological potential for wearable health monitoring and touchless control panels. The demonstrated behavior is unparalleled in literature, surpassing other state of the art solutions in terms of sensor response and recovery times, ease of manufacture, substrate compatibility, transparency and scale-up potential.

**Acknowledgements:** This work was financially supported by the Ministry of Education, Science and Technological Development of the Republic of Serbia (Grant No. 451-03-68/2020-14/200026). The authors acknowledge funding provided by the Institute of Physics Belgrade, through the grant by the Ministry of Education, Science, and Technological Development of the Republic of Serbia. This research was supported by the Science Fund of the Republic of Serbia, PROMIS, #6057070, Gramulsen.

**References**

[1] X. Liu, S. Cheng, H. Liu, S. Hu, D. Zhang, H. Ning, A survey on gas sensing technology, Sensors (Switzerland). 12 (2012) 9635–9665. https://doi.org/10.3390/s120709635.

[2] T.A. Blank, L.P. Eksperiandova, K.N. Belikov, Recent trends of ceramic humidity sensors development: A review, Sensors Actuators, B Chem. 228 (2016) 416–442. https://doi.org/10.1016/j.snb.2016.01.015.

[3] A. Manolis, The diagnostic potential of breath analysis, Clin. Chem. 29 (1983) 5–15.

[4] T. Earthrowl-Gould, B. Jones, M.R. Miller, Chest and abdominal surface motion measurement for continuous monitoring of respiratory function, Proc. Inst. Mech. Eng. Part H J. Eng. Med. 215 (2001) 515–520. https://doi.org/10.1243/0954411011536028.

[5] M. Zirkl, A. Sawatdee, U. Helbig, M. Krause, G. Scheipl, E. Kraker, P.A. Ersman, D. Nilsson, D. Platt, P. Bodö, S. Bauer, G. Domann, B. Stadlober, An all-printed ferroelectric active matrix sensor network based on only five functional materials forming a touchless control interface, Adv. Mater. 23 (2011) 2069–2074. https://doi.org/10.1002/adma.201100054.

[6] J. Feng, L. Peng, C. Wu, X. Sun, S. Hu, C. Lin, J. Dai, J. Yang, Y. Xie, Giant Moisture Responsiveness of $VS_2$ Ultrathin Nanosheets for Novel Touchless Positioning Interface, Adv. Mater. 24 (2012) 1969–1974. https://doi.org/10.1002/adma.201104681.

[7] K. Nguyen, C.M. Hung, T.M. Ngoc, D.T. Thanh Le, D.H. Nguyen, D. Nguyen Van, H. Nguyen Van, Low-temperature prototype hydrogen sensors using Pd-decorated SnO2 nanowires for exhaled breath applications, Sensors Actuators, B Chem. 253 (2017) 156–163. https://doi.org/10.1016/j.snb.2017.06.141.

[8] R. Yoo, Y. Park, H. Jung, H.J. Rim, S. Cho, H.S. Lee, W. Lee, Acetone-sensing properties of doped ZnO nanoparticles for breath-analyzer applications, J. Alloys Compd. 803 (2019) 135–144. https://doi.org/10.1016/j.jallcom.2019.06.254.

[9] S. Papamatthaiou, D.P. Argyropoulos, F. Farmakis, A. Masurkar, K. Alexandrou, I. Kymissis, N. Georgoulas, The Effect of Thermal Reduction and Film Thickness on fast Response Transparent Graphene Oxide Humidity Sensors, Procedia Eng. 168 (2016) 301–304. https://doi.org/10.1016/j.proeng.2016.11.201.

[10] A.D. Smith, K. Elgammal, F. Niklaus, A. Delin, A.C. Fischer, S. Vaziri, F. Forsberg, M. Råsander, H. Hugosson, L. Bergqvist, S. Schröder, S. Kataria, M. Östling, M.C. Lemme, Resistive graphene humidity sensors with rapid and direct electrical readout, Nanoscale. 7 (2015) 19099–19109. https://doi.org/10.1039/c5nr06038a.

[11]   X. Fan, K. Elgammal, A.D. Smith, M. Östling, A. Delin, M.C. Lemme, F. Niklaus, Humidity and CO2 gas sensing properties of double-layer graphene, Carbon N. Y. 127 (2018) 576–587. https://doi.org/10.1016/j.carbon.2017.11.038.

[12]   A.P. Taylor, L.F. Velásquez-García, Electrospray-printed nanostructured graphene oxide gas sensors, Nanotechnology. 26 (2015) 505301. https://doi.org/10.1088/0957-4484/26/50/505301.

[13]   S. Xiao, J. Nie, R. Tan, X. Duan, J. Ma, Q. Li, T. Wang, Fast-response ionogel humidity sensor for real-time monitoring of breathing rate, Mater. Chem. Front. 3 (2019) 484–491. https://doi.org/10.1039/c8qm00596f.

[14]   S. Borini, R. White, D. Wei, M. Astley, S. Haque, E. Spigone, N. Harris, J. Kivioja, T. Ryhänen, Ultrafast graphene oxide humidity sensors, ACS Nano. 7 (2013) 11166–11173. https://doi.org/10.1021/nn404889b.

[15]   B. Jiang, Z. Bi, Z. Hao, Q. Yuan, D. Feng, K. Zhou, L. Zhang, X. Gan, J. Zhao, Graphene oxide-deposited tilted fiber grating for ultrafast humidity sensing and human breath monitoring, Sensors Actuators, B Chem. 293 (2019) 336–341. https://doi.org/10.1016/j.snb.2019.05.024.

[16]   T. Tomašević-Ilić, Đ. Jovanović, I. Popov, R. Fandan, J. Pedrós, M. Spasenović, R. Gajić, Reducing sheet resistance of self-assembled transparent graphene films by defect patching and doping with UV/ozone treatment, Appl. Surf. Sci. 458 (2018) 446–453. https://doi.org/10.1016/J.APSUSC.2018.07.111.

[17]   C. Mackin, A. Fasoli, M. Xue, Y. Lin, A. Adebiyi, L. Bozano, T. Palacios, Chemical sensor systems based on 2D and thin film materials, 2D Mater. (2020). https://doi.org/10.1088/2053-1583/ab6e88.

[18]   A. Matković, I. Milošević, M. Milićević, T. Tomašević-Ilić, J. Pešić, M. Musić, M. Spasenović, D. Jovanović, B. Vasić, C. Deeks, R. Panajotović, M.R. Belić, R. Gajić, Enhanced sheet conductivity of Langmuir-Blodgett assembled graphene thin films by chemical doping, 2D Mater. 3 (2016). https://doi.org/10.1088/2053-1583/3/1/015002.

[19]   T. Tomašević-Ilić, J. Pešić, I. Milošević, J. Vujin, A. Matković, M. Spasenović, R. Gajić, Transparent and conductive films from liquid phase exfoliated graphene, Opt. Quantum Electron. 48 (2016) 1–7. https://doi.org/10.1007/s11082-016-0591-1.

[20]   H. Kim, C. Mattevi, H.J. Kim, A. Mittal, K.A. Mkhoyan, R.E. Riman, M. Chhowalla, Optoelectronic properties of graphene thin films deposited by a Langmuir-Blodgett assembly, Nanoscale. 5 (2013) 12365–12374. https://doi.org/10.1039/C3NR02907G.

[21]   M. Sarajlić, M. Frantlović, M.M. Smiljanić, M. Rašljić, K. Cvetanović-Zobenica, Ž. Lazić, D. Vasiljević-Radović, Thin-film four-resistor temperature sensor for measurements in air, Meas. Sci. Technol. 30 (2019) 115102.

https://doi.org/10.1088/1361-6501/AB326C.

[22] X. Li, G. Zhang, X. Bai, X. Sun, X. Wang, E. Wang, H. Dai, Highly conducting graphene sheets and Langmuir-Blodgett films, Nat. Nanotechnol. 3 (2008) 538–542. https://doi.org/10.1038/nnano.2008.210.

[23] W.M.R. Divigalpitiya, R.F. Frindt, S.R. Morrison, Effect of humidity on spread NbS2films, J. Phys. D. Appl. Phys. 23 (1990) 966–970. https://doi.org/10.1088/0022-3727/23/7/035.

[24] F. Yavari, C. Kritzinger, C. Gaire, L. Song, H. Gulapalli, T. Borca-Tasciuc, P.M. Ajayan, N. Koratkar, Tunable bandgap in graphene by the controlled adsorption of water molecules, Small. 6 (2010) 2535–2538. https://doi.org/10.1002/smll.201001384.

[25] E. Massera, V. La Ferrara, M. Miglietta, T. Polichetti, I. Nasti, G. Di Francia, Gas sensors based on graphene. Comparison of two different fabrication approaches, Chim. Oggi/Chemistry Today. 30 (2012) 29–31.

[26] M. Sarajlić, Z. Đurić, V. Jović, S. Petrović, D. Đorđević, Detection limit for an adsorption-based mercury sensor, Microelectron. Eng. 103 (2013) 118–122. https://doi.org/10.1016/j.mee.2012.10.009.

[27] Y. Pang, J. Jian, T. Tu, Z. Yang, J. Ling, Y. Li, X. Wang, Y. Qiao, H. Tian, Y. Yang, T.L. Ren, Wearable humidity sensor based on porous graphene network for respiration monitoring, Biosens. Bioelectron. 116 (2018) 123–129. https://doi.org/10.1016/j.bios.2018.05.038.

[28] H. Cheng, Y. Huang, L. Qu, Q. Cheng, G. Shi, L. Jiang, Flexible in-plane graphene oxide moisture-electric converter for touchless interactive panel, Nano Energy. 45 (2018) 37–43. https://doi.org/10.1016/j.nanoen.2017.12.033.

[29] J. An, T.S.D. Le, Y. Huang, Z. Zhan, Y. Li, L. Zheng, W. Huang, G. Sun, Y.J. Kim, All-Graphene-Based Highly Flexible Noncontact Electronic Skin, ACS Appl. Mater. Interfaces. 9 (2017) 44593–44601. https://doi.org/10.1021/acsami.7b13701.

[30] B.H. Wee, W.H. Khoh, A.K. Sarker, C.H. Lee, J.D. Hong, A high-performance moisture sensor based on ultralarge graphene oxide, Nanoscale. 7 (2015) 17805–17811. https://doi.org/10.1039/c5nr05726d.

# Supporting information

## Ultrafast humidity sensor based on liquid phase exfoliated graphene

S.Andrić[1], T. Tomašević-Ilić[2], M. V. Bošković[1], M. Sarajlić[1], D. Vasiljević-Radović[1] M. M. Smiljanić[1] and M. Spasenović[1]

[1]Center for Microelectronic Technologies, Institute of Chemistry, Technology and Metallurgy, University of Belgrade, Njegoševa 12, 11000 Belgrade, Serbia

[2] Institute of Physics Belgrade, University of Belgrade, Pregrevica 118, 11080 Belgrade, Serbia

Corresponding author: spasenovic@nanosys.ihtm.bg.ac.rs

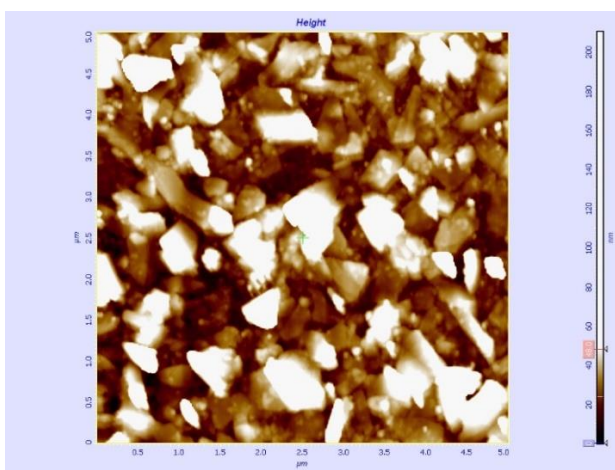

**Figure S1**. AFM characterization of graphene film on Si/SiO$_2$, with nanoplatelets visible.

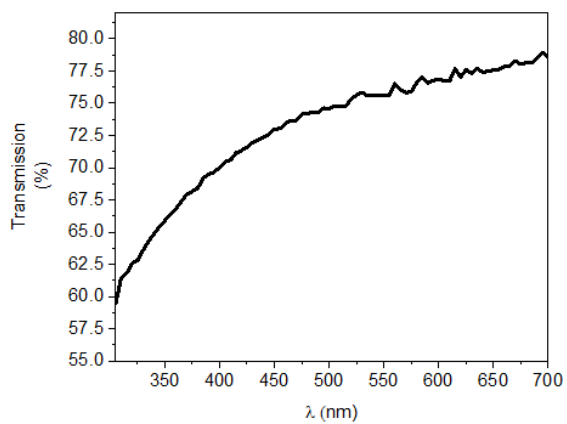

**Figure S2**. UV-VIS transmission spectrum of graphene film on PET. A transmission of 77% at a wavelength of 660 nm indicates average film thickness of 10 layers (3.4 nm) [1].

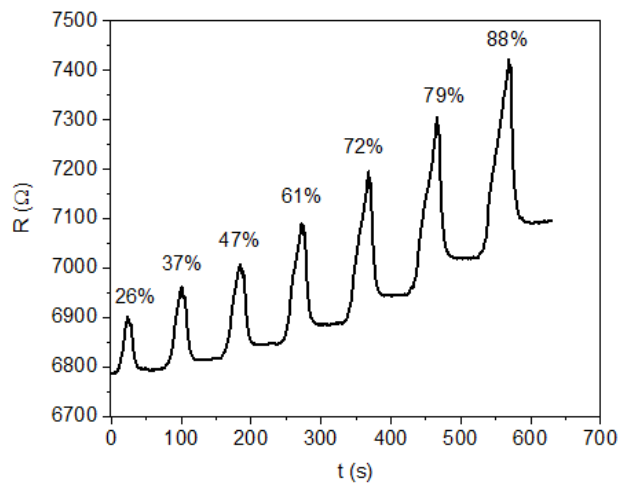

**Figure S3**. Raw data of sensor response to repeated stepwise increase of relative humidity.

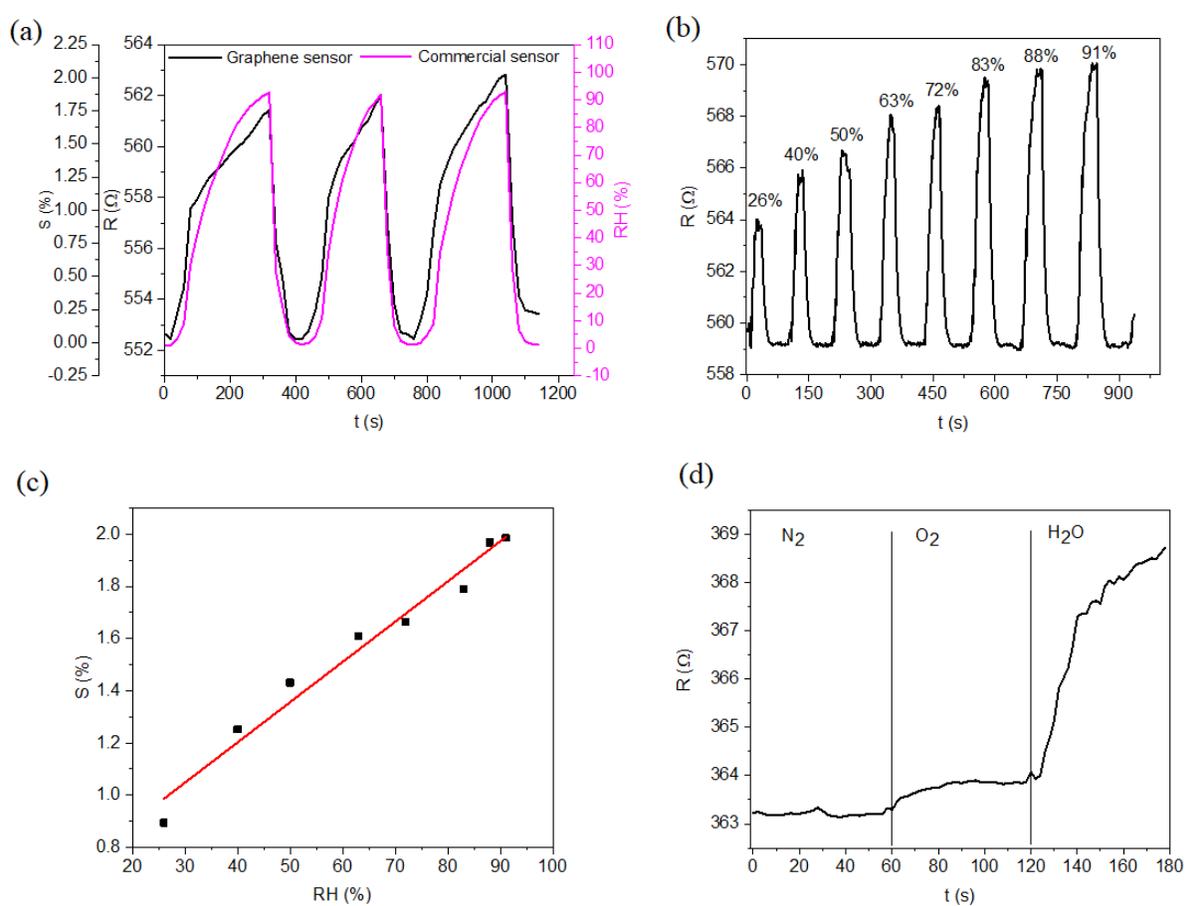

**Figure S4**. (a) Graphene sensor on Si/SiO$_2$ (black) and commercial sensor response to changing RH from ~ 0 % to 96%. (b) Graphene sensor on Si/SiO$_2$ response to stepwise increase of relative humidity, from ~ 0 % to 91%. (c) Graphene sensor on Si/SiO$_2$ response as function of maximum humidity. (d) Graphene sensor on on Si/SiO$_2$ selectivity to nitrogen, oxygen and water molecules in the vacuum chamber for 60 seconds each.

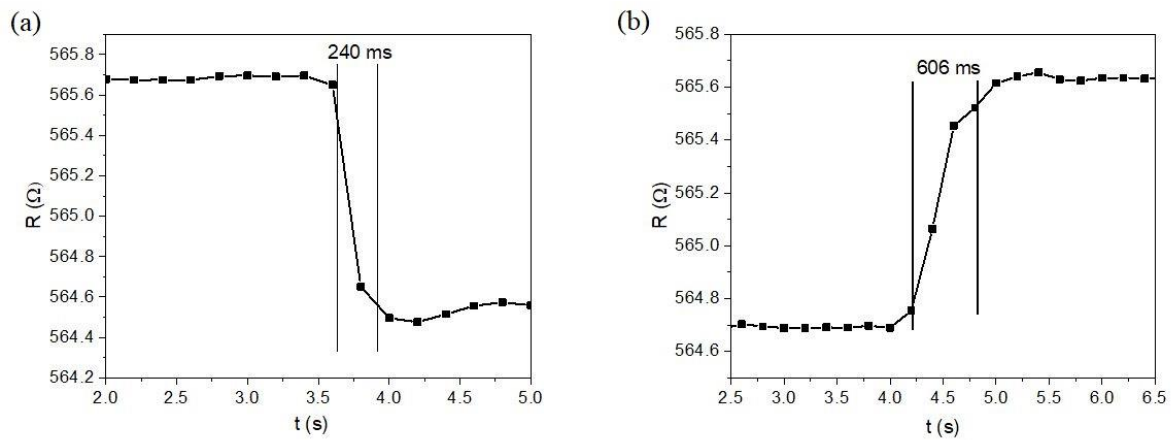

**Figure S5**. Graphene sensor on Si/SiO$_2$ response at room conditions to exposure to gas flow from a nitrogen gun. (b) Response time. (c) Recovery time. The thresholds for measuring response and recovery times were 10% and 90% change in resistance.

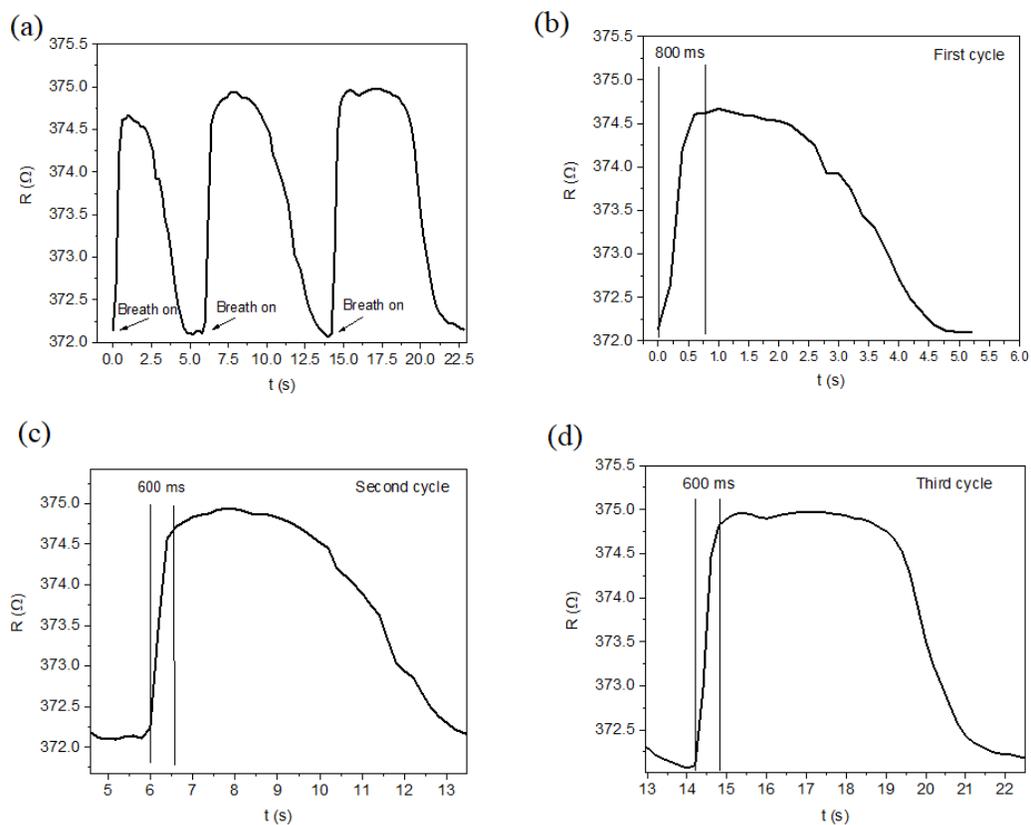

**Figure S6**. (a) Breath monitoring with graphene on Si/SiO$_2$. (b) Response time of the first cycle. (c) Response time of the second cycles. (d) Response time of the third cycle.

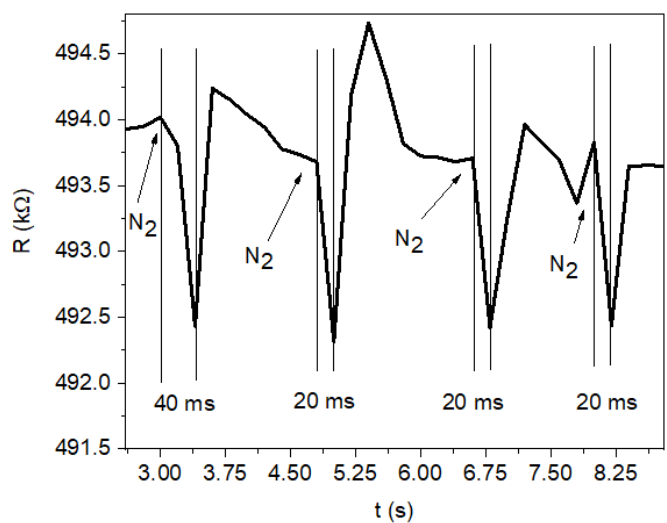

**Figure S7**. Resistance response of the graphene sensor on PET at room conditions while exposed to nitrogen gun in four cycles, with response time indicated.

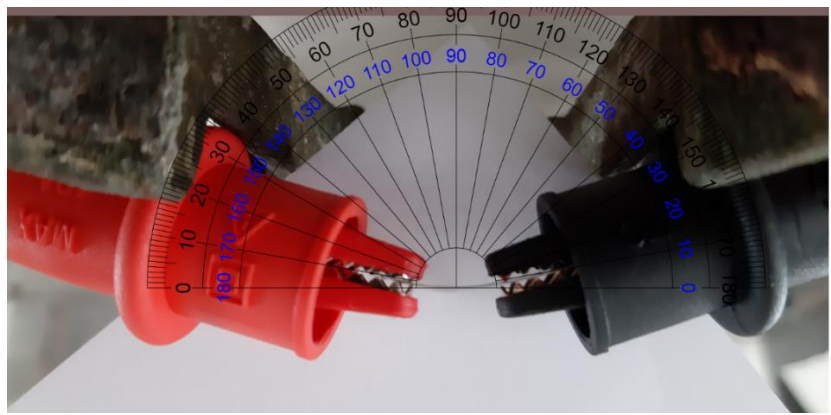

**Figure S8.** PET substrate flexed at an angle of 10 degrees relative to the horizon.

**Supplementary material references**

[1]　　Bonaccorso F, Sun Z, Hasan T and Ferrari A C 2010 Graphene photonics and optoelectronics *Nature Photonics* **4** 611–22